\DeclareUnicodeCharacter{02BC}{'}
\documentclass[aip,apl,reprint]{revtex4-1}

\usepackage{graphicx}
\usepackage{dcolumn}
\usepackage{bm}

\usepackage[utf8]{inputenc}
\usepackage[T1]{fontenc}
\usepackage{mathptmx}
\usepackage{etoolbox}

\usepackage{graphicx}
\usepackage[dvipsnames]{xcolor}
\usepackage{amsmath}
\usepackage{amssymb,gensymb}
\usepackage{lineno}
\usepackage{subfigure}
\usepackage[normalem]{ulem}
\graphicspath{{./figures/}}

\begin{document}

\preprint{AIP/123-QED}

\title{Seismic Isolation of Optical Tables Using Piezo Actuators}

\author{Tailong Wang}
\email{tl.wang.au@outlook.com}
\affiliation{
\parbox{0.9\textwidth}{School of Electrical and Electronic Engineering, Wuhan Polytechnic University, Wuhan 430023, China}
}
\affiliation{OzGrav, University of Western Australia, Crawley, Western Australia 6009, Australia}

\author{Carl Blair}
\email{carl.blair@uwa.edu.au}
\affiliation{OzGrav, University of Western Australia, Crawley, Western Australia 6009, Australia}

\author{Ammar Al-Jodah}
\affiliation{OzGrav, University of Western Australia, Crawley, Western Australia 6009, Australia}

\author{John Winterflood}
\affiliation{OzGrav, University of Western Australia, Crawley, Western Australia 6009, Australia}

\author{Jian Liu}
\affiliation{OzGrav, University of Western Australia, Crawley, Western Australia 6009, Australia}

\author{Alexander Adams}
\affiliation{OzGrav, University of Western Australia, Crawley, Western Australia 6009, Australia}

\author{Aaron Goodwin-Jones}
\affiliation{OzGrav, University of Western Australia, Crawley, Western Australia 6009, Australia}
\affiliation{California Institute of Technology, Pasadena, California 91125, USA}

\author{Chunnong Zhao}
\affiliation{OzGrav, University of Western Australia, Crawley, Western Australia 6009, Australia}

\author{Li Ju}
\affiliation{OzGrav, University of Western Australia, Crawley, Western Australia 6009, Australia}

\date{\today}

\begin{abstract}
Seismic isolation is crucial for gravitational wave detectors as it minimizes ground vibrations, enabling the detection of faint gravitational wave signals.
 An active seismic isolation platform for precision measurement experiments is described.  
 The table features piezo actuation along five degrees of freedom: three translational actuations and two tip-tilt degrees of freedom along the horizontal axes. It is stiff in rotation about the vertical axes.
 A seismometer is used to sense table motion.  
 Piezo actuators are used to suppress seismic noise with feedback control bandwidth of 0.3 to 3\,Hz. 
 Suppression levels ranging from 21 to 36\,dB of seismic noise within the frequency range of 0.5 to 1.3\,Hz are demonstrated, as measured by a witness seismometer on the table, with the suppression direction along the axis of the longitudinal translation of the suspended mirror on the table. 
The suppression results in 1\,$\mathrm{\mathrm{nm/\sqrt{Hz}}}$ residual horizontal motion at 1\,Hz.
 Limitations such as tilt-to-translation coupling that prevent actuation over the desired range of 0.03 to 3\,Hz are discussed.
\end{abstract}

\maketitle

\section{Introduction}
  
Advanced gravitational wave detectors (GWDs), such as Advanced LIGO (aLIGO)\cite{ligo15}, Advanced Virgo\cite{acernese2015}, and KAGRA\cite{aso13}, require highly effective vibration isolation systems. These systems attenuate seismic noise by nine orders of magnitude at frequencies around 100 Hz, where gravitational waves (GW) are detected.
Seismic isolation requirements out of the GWD frequency band are less well-defined.  
Less effective seismic isolation below 30\,Hz, through many complicated channels, has not only limited the GW signal detection at lower frequencies, but also reduced detector duty cycle at LIGO\cite{marty16,PhysRevD.102.062003,Nguyen_2021}.  

We aim to reduce microseismic noise and seismic excitation of suspension resonances at the Gingin High Optical Power Facility (HOPF).
In this paper, we present the upgraded optical bench, which functions as a five-degree-of-freedom piezo active isolator for seismic isolation. 
The advantage of this system is the simplicity of the design.
It is able to actuate heavy payloads with a compact and easily manufactured actuator.
Optics are suspended by pendulum with a free-swinging resonant frequency around 0.5\,Hz to suppress seismic noise at high frequencies.
To mitigate the effects of microseismic noise and seismic excitation of suspension resonances, the target frequency band for our active isolator is from 0.03\,Hz to 3\,Hz. Herein we label it the active pre-isolator.
The plan is to implement the active pre-isolator across all auxiliary optic tables, and use it as a pre-isolation stage for the University of Western Australia (UWA) Compact Vibration Isolation System (CVIS)\cite{EJChin_2006,barr09} primarily to mitigate tilt coupling to the CVIS.
The piezo pre-isolator is currently installed in two isolation tables for single pendulum stage suspended optics at the HOPF.

Active pre-isolation systems are used in the Advanced LIGO\cite{Matichard_2015} and VIRGO\cite{BLOM2013466} detectors as well as the AEI 10m-prototype\cite{Wanner_2012} (AEI-SAS).
The AEI-SAS uses a combination of low-frequency passive isolation and active control to achieve an impressive 40\,dB suppression of horizontal table motion around 1\,Hz.  
The frequency range between 0.08\,Hz and 0.7\,Hz relies on the L4C geophone and is limited by self-noise below 0.13\,Hz\cite{Kirchhoff:2022ukf}.  
In this frequency range, AEI-SAS provides a maximum of 30\,dB translation suppression.  At 1\,Hz the AEI-SAS achieves 0.5\,$\mathrm{nm/\sqrt{Hz}}$.

Advanced LIGO uses three different active pre-isolation systems\cite{Matichard_2015}: Hydraulic External Pre-Isolator (HEPI)\cite{Wen_2014}, Basic Symmetric Chambers - Internal Seismic Isolation (BSC-ISI) and Horizontal Access Module - Internal Seismic Isolation (HAM-ISI)\cite{kisselenhanced}. 
The HEPI system supports subsequent isolation stages and it is designed to provide one stage of very low frequency active isolation for the BSC or HAM chambers. They are currently used for alignment and tidal drift between chambers\cite{Matichard_2015}.
The ISI is supported by the HEPI system and is used to provide a second stage of pre-isolation\cite{Aston_2012}. 
The performance of the combination of HEPI and BSC-ISI pre-isolation shows 60\,dB suppression between 0.1\,Hz and 1\,Hz. 
It achieves nearly 60\,dB suppression between 1\,Hz and 10\,Hz.  
The measurement in the relevant frequency between 0.9\,Hz and 5\,Hz is limited by GS13 (geophone) sensor noise\cite{Matichard_2015}. 
The HAM-ISI system is similar to the Gingin system as it supports the auxiliary optics\cite{aasi2015advanced}. It is also pre-isolated by the HEPI, and the combined performance shows 60\,dB suppression at 1\,Hz and roughly 10\,dB suppression at 0.25\,Hz. Detailed performance analysis is in the paper\cite{Matichard_2015}.
Both the HAM-ISI and BSC-ISI achieve residual motion (measured by an in loop GS13) of 0.01\,$\mathrm{nm/\sqrt{Hz}}$ at 1\,Hz.
As a historical note, prior to HEPI, LIGO Livingston used a piezo pre-isolation system called PEPI \cite{RAbbott_2004}.  This system only worked in one degree of freedom and achieved up to 20\,dB suppression of resonant motion around 0.75\,Hz and 1.3\,Hz as witnessed by LIGO optical cavities.

The External Injection Bench Seismic Attenuation System (EIB-SAS) on Advanced Virgo is used to damp the resonance of the EIB support structure \cite{BLOM2013466}. 
It uses both active and passive isolation similar to AEI-SAS. Located in air and carrying auxiliary optics, this system demonstrates good active damping performance, achieving 30-70 dB horizontal suppression between 2 Hz and 400 Hz. 
Compared to the active pre-isolation system described here, this system relies more heavily on passive isolation. 
The EIB-SAS does not require suppression in the micro-seismic frequency band \cite{BLOM2015641}, however some suppression is achieved in the translation degrees of freedom. 
The EIB-SAS achieved 10\,$\mathrm{nm/\sqrt{Hz}}$ at 1\,Hz, however this is not in the target frequency band, it achieves about 0.1\,$\mathrm{nm/\sqrt{Hz}}$ at 10\,Hz \cite{BlomThesis}.

As a commercial comparison, the THORLAB active pre-isolation table\cite{thorlabs2018optical} achieves isolation above 2\,Hz, reaching -24\,dB at 5\,Hz and -30\,dB at 10\,Hz horizontally. However, it lacks attenuation below its horizontal resonant frequency of approximately 1\,Hz. 
In UWA basement labs residual motion on these tables is a few \,$\mathrm{nm/\sqrt{Hz}}$ at 10\,Hz.
The active pre-isolator described in this paper provides competitive performance in the microseism frequency range with 36\,dB suppression and a residual motion of 4\,$\mathrm{nm/\sqrt{Hz}}$ at 0.5\,Hz, similar to the ISI at this frequency.  While it does not fulfill requirements of tables presented above, it does provide a relatively simple solution that may be incorporated in other systems to provide isolation in this frequency range.

In section \ref{sec:descr} the apparatus is described.  In section \ref{sec:exp} the system has been experimentally demonstrated to achieve suppression levels ranging from 36 to 21\,dB between 0.5 and 1.3\,Hz. In section \ref{sec:performance} we describe performance limitations, future plans and summarize system performance.

\section{Description of Apparatus}\label{sec:descr}

\begin{figure}[h!]
	\centering	\includegraphics[width=0.45\textwidth]{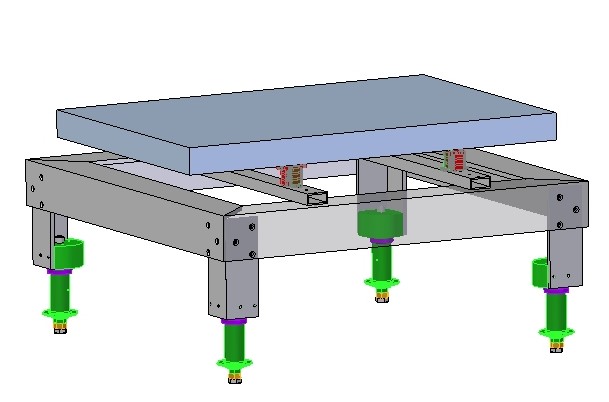}\\
	\caption{Schematic CAD model of active pre-isolation table.  The green parts are the feet with vertical actuation.  The red units are the horizontal actuators.}
	\label{fig:table_schem}
\end{figure}

\begin{figure*}[ht]
	\centering	\includegraphics[width=0.8\textwidth]{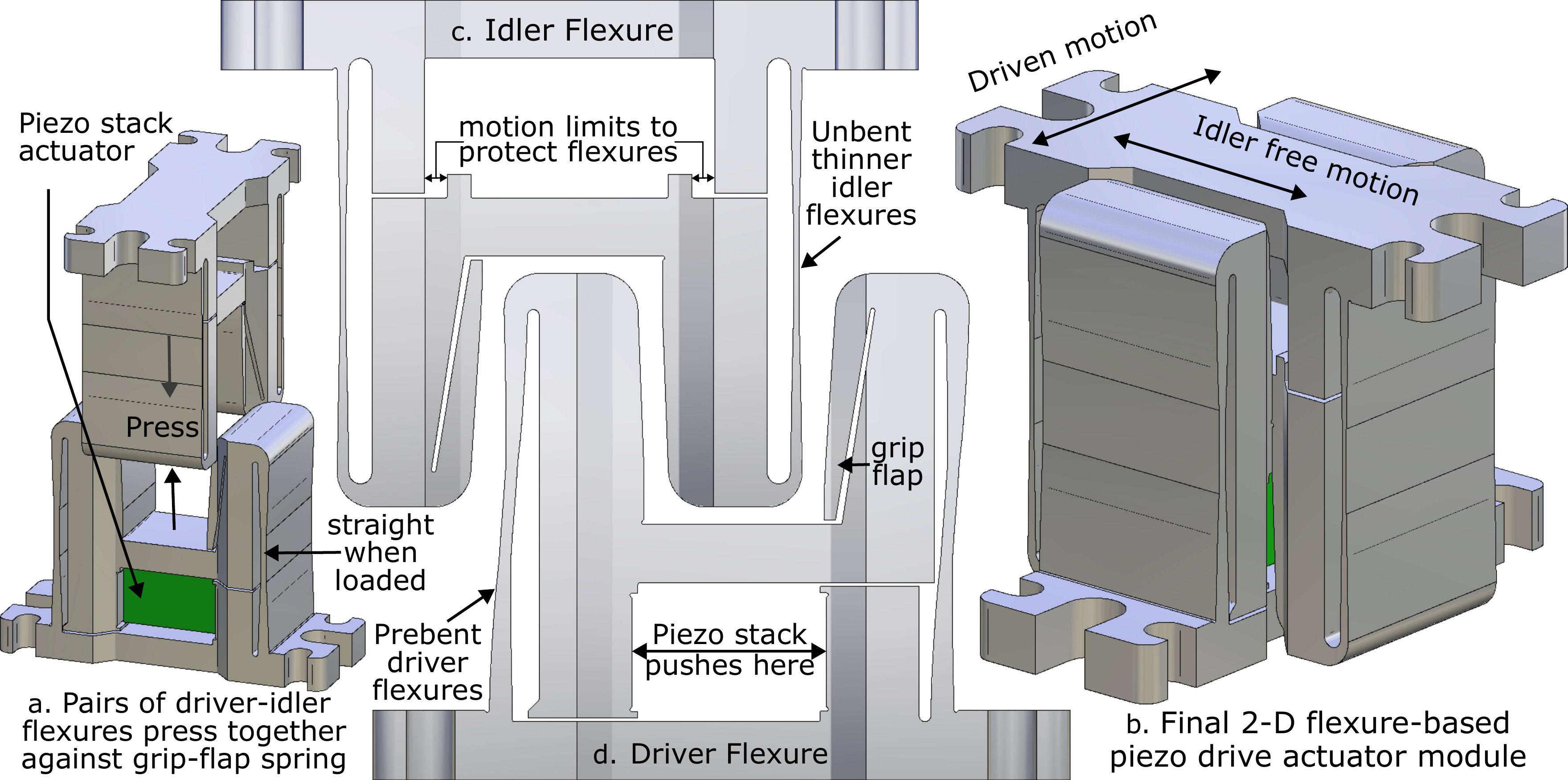}\\
	\caption{Horizontal piezo actuator stage. a) Exploded view shows the pair of flexures is pressed together to form the actuator.  b) The assembled actuator with free and driven degree of freedom indicated. c) The idler (top) flexure profile. d) The driver (base) flexure profile.
 }
	\label{fig:hor_act}
\end{figure*}

The active pre-isolation tables design shown in Figure \ref{fig:table_schem} is as follows.  
A frame is mounted on feet with a screw adjustment for frame level alignment.
Four piezo actuators are incorporated into the foot screw design to provide three degrees of freedom actuation in vertical, pitch and roll from the test mass perspective.
The optical table sits upon four horizontal piezo actuators that are mounted to the frame.
The horizontal piezo actuators are mounted in a novel structure fabricated by Electrical Discharge Machine (EDM), which allows free movement in the degree of freedom that is not driven by the piezo actuators as shown in Figure \ref{fig:hor_act}.
In this way two of the four actuate in one horizontal degree of freedom and two actuate in the orthogonal degree of freedom.
Generally one of these degrees of freedom is aligned with the optic axis.

\begin{figure}[h!]
	\centering	\includegraphics[width=0.45\textwidth]{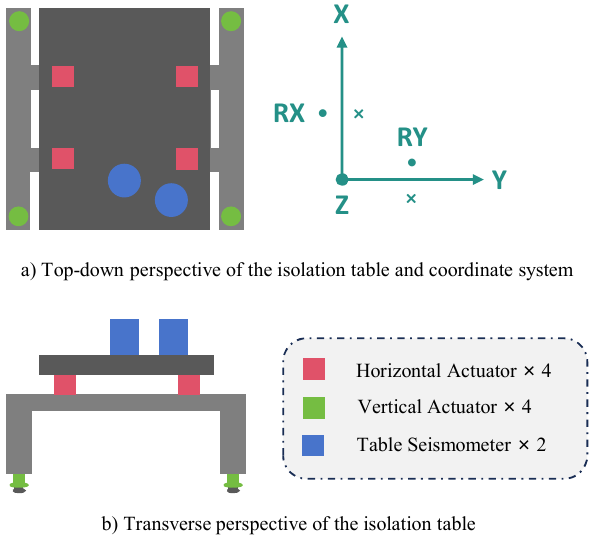}\\
	\caption{Simplified illustration of active pre-isolation table: (a) Top-down perspective illustrating the arrangement of sensors and actuators, along with the coordinate system representing the degrees of freedom controlled by the actuators. (b) Transverse perspective depicting the vertical alignment of sensors and actuators.}
	\label{fig:table_view}
\end{figure}

A simplified schematic of the active pre-isolation table is presented in Figure \ref{fig:table_view}. Two Nanometrics, Trillium Compact T120 seismometers acting as sensors on the table are used for control and out of loop witness of residual motion. An additional T120 seismometer on the ground is used to estimate table seismic noise suppression.  The coordinate system depicted in Figure \ref{fig:table_view} corresponds to a top-down view, where the seismometers measure motion along the X, Y, and Z axes. From the perspective of the active pre-isolation table, the vertical actuators provide degrees of freedom in the Z, RX (rotational motion about the X-axis), and RY (rotational motion about the Y-axis) directions, while the horizontal actuators control motion in the X and Y directions. 

\begin{figure}[h!]
	\centering	\includegraphics[width=0.45\textwidth]{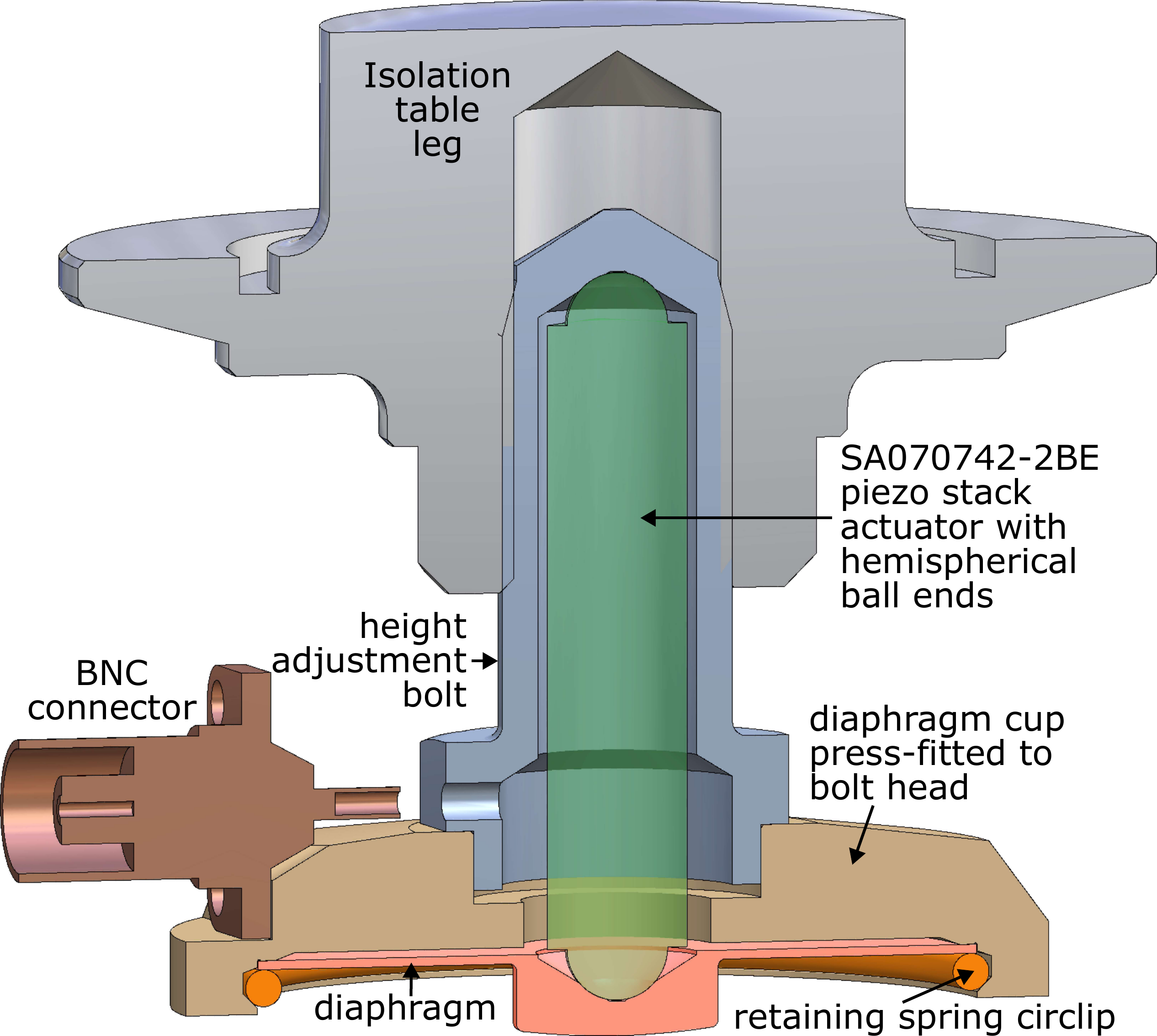}\\
	\caption{The vertical actuator housed in the height adjustment foot of the table.  The Isolation table leg supports the table inside the vacuum tank and forms the vacuum flange to flexible bellows (not shown).  The threaded height adjustment bolt (threads not shown) is screwed into the isolation table leg.  The piezo stack is inserted into the bolt.  The diaphragm cup attaches to the bolt with a circular interference fit.  The diaphragm holds the piezo in place and is held into the cup with a circlip.  A BNC connector is used to provide voltage to the piezo via wires that are not shown.  A pin (not shown) prevents rotation between bolt, cup and diaphragm.}
	\label{fig:vert_act}
\end{figure}

The vertical actuator is composed of a piezo stack actuator (SA070742, Piezodrive) mounted inside a height adjustment M16, 1.5\,mm pitch, 45\,mm long, hardened steel  bolt forming the foot of the table as shown in Figure \ref{fig:vert_act}. 
The bolt allows for coarse table height adjustment and is bored hollow to make room for the piezo. 
The upper end of the piezo pushes on the bolt, while the lower end pushes on the floor via a flexible diaphragm. 
The head of the bolt is machined to mate with a cup which houses a diaphragm to constrain the motion of the piezo so that it follows a straight vertical line and prevents it from being affected by any shear forces if the foot is dragged across the floor. 
A pin (not shown) that press fits off center into the bolt, cup, and diaphragm was added to ensure the diaphragm turns with the bolt and cup during height adjustment to prevent torque being applied to the piezo. 
Torque is also minimised by lightly greasing the ball-ends of the piezo for assembly. 
Each piezo supports a load of 180 kg (for 720 kg total) with a 70\,$\mu$m actuation range.

The horizontal actuators are constructed of two similar single-axis monolithic flexures, 'driver' and 'idler', pressed together to form a 2-axis translation stage as shown in Figure \ref{fig:hor_act}. 

The driver flexure mechanism has a bend machined in such that when a piezo (10x10x18 mm) is fitted, the flexures become straight, and the resulting stress provides the standard piezo preload requirement. 
The idler flexure mechanism has thinner straight flexures and provides simple single-axis low friction load support for the orthogonal direction. 
The optical table is mounted on 4 of these 2-axis translation stages with one diagonal pair oriented so that the piezo drive acts in the X direction and free-wheels in the Y direction, while the other diagonal pair are oriented orthogonal to provide stiff drive in the Y direction and free-wheeling in the X. In this way piezo-driven pure X-Y translation is obtained, with all other degrees of freedom as stiff as the flexures in their non-S-bend directions provides.

\begin{table}[h]
	\caption{Table Properties}
	\centering
	\renewcommand{\arraystretch}{1}
	\begin{tabular}{c | c } 
		\hline
		Parameters & Value \\
		\hline
		Mass of Table, $m_T$& 20 kg \\
		Mass of Frame, $m_F$& 10 kg \\
		Vertical Actuation Range, $l_v$  & 70 $\mu$m \\
		Horizontal Actuation Range, $l_h$  & 40 $\mu$m \\
        Tilt Actuation Range, $\phi$  & 80 urad \\
		First resonant frequency, $\omega_m1$ & $\sim$ 15 Hz\\
		Seismometers (Nanometrics), &  3x T120s\\
		Suspension height, $h_S$& 650 mm \\
		Mirror mass weight, $m_S$ & 880 g \\
		\hline
	\end{tabular}
	\label{tab:gingincavityparameters}
\end{table}

The table was simulated in ANSYS \cite{desalvo1985ansys} eigenfrequency analysis. The simulation shows the lowest internal mode frequency is 45\,Hz.  However, in practice the lowest mode frequency is about 15\,Hz and depends on the adjustment of the 4 feet of the table.
The lower frequency internal resonance of the table limits the control band.
As a result, we set a nominal target of 40\,dB suppression from 0.03-3\,Hz.

\section{Experimental Demonstration}\label{sec:exp}

\begin{figure}[h!]
	\centering
	\includegraphics[width=0.45\textwidth]{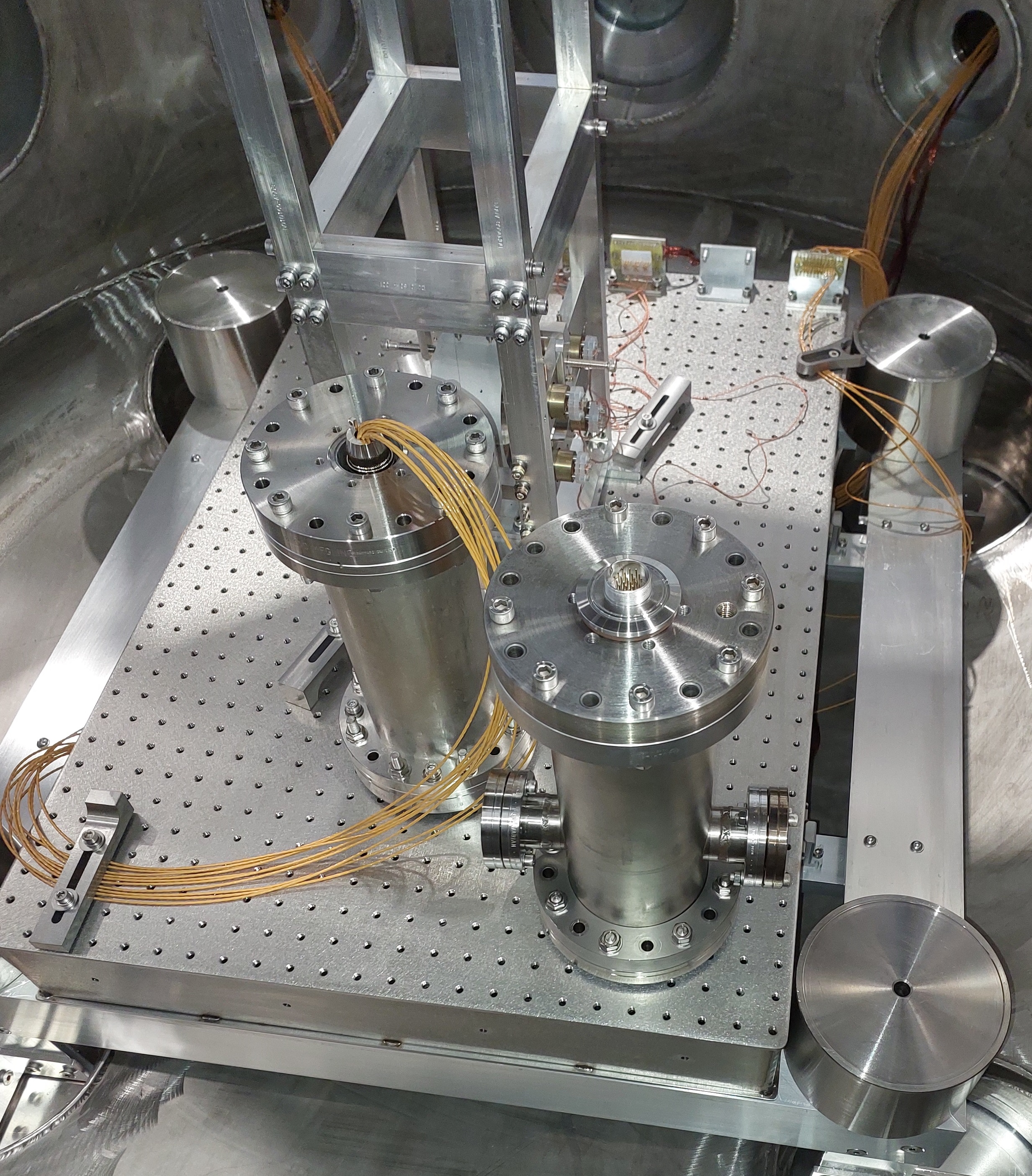}\\
	\caption{Photograph of one of the isolated tables during construction.  The 2 seismometers are inside the cylindrical cans.  There is a signal suspended mirror in the center of the table.}
	\label{fig:tablephoto}
\end{figure}

The Gingin facility is used as a technology demonstrator for gravitational wave detector technology.  The facility currently houses, 2 long suspended optical cavity experiments, one 74\,m cavity with fused silica optics\cite{PhysRevD.91.092001} and another with a 2\,$\mu$m laser and silicon optics\cite{10.1063/5.0136869}.  
It is also the location of a seismic array\cite{Satari_2022} that is being evaluated for vibration isolation improvements and Newtonian noise subtraction.  
The experimental demonstration reported here is on two tables for auxiliary optics for the 2\,$\mu$m cavity experiments.  These 2 tables are separated by 7\,m.  
We focus on the optic axis (Y direction) of the input-optic-table, which currently supports a single suspended optic.  
Results presented here show control in just this axis.
A photo of the table is shown in Figure~\ref{fig:tablephoto}.

\begin{figure}[h!]
	\centering
	\includegraphics[width=0.48\textwidth]{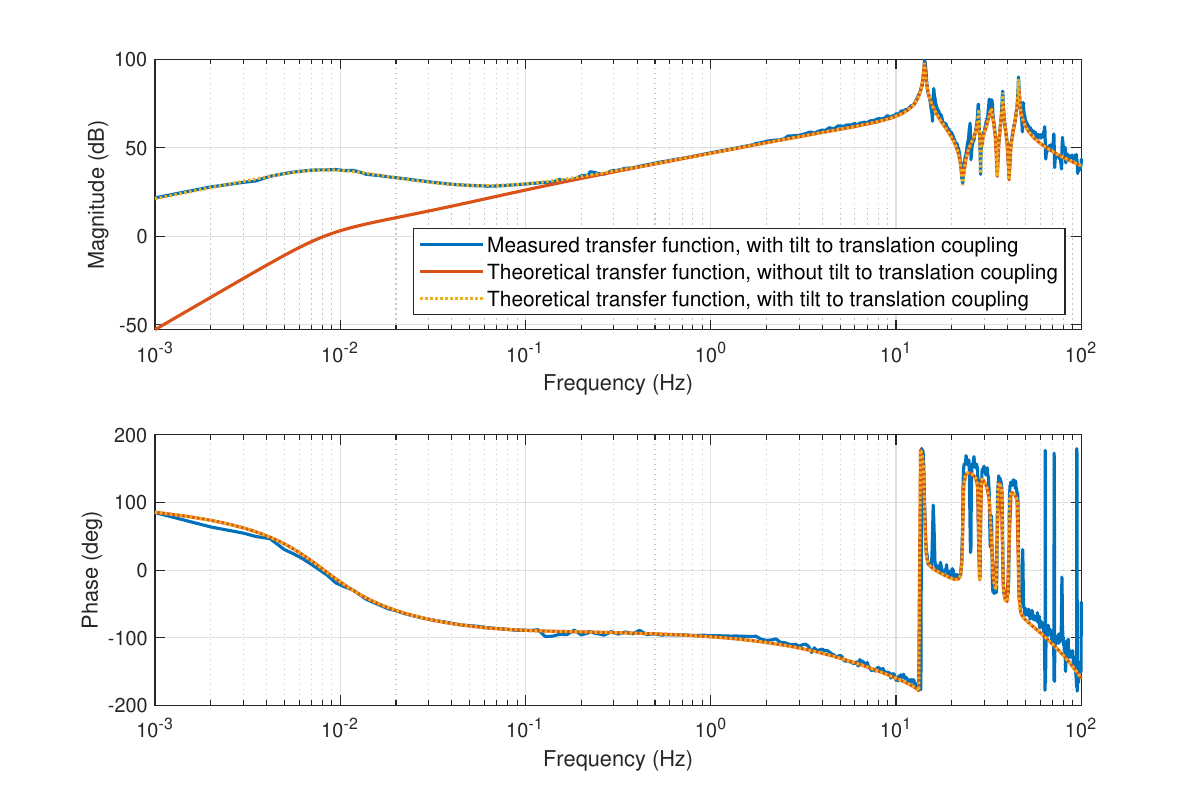}\\
    \includegraphics[width=0.42\textwidth]{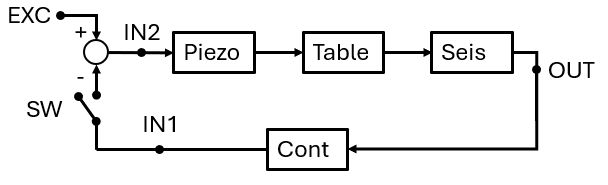}
	\caption{Top 2 panels - Bode plot of the table transfer function measured as OUT/EXC with SW open in the lower panel control diagram. The blue curve represents the measured transfer function, while the red and yellow curves show the theoretical transfer functions without and with tilt to translation coupling, respectively (including fitted resonances). The discrepancy below 100\,mHz is explained by table tilt coupling. Control diagram shows Excitation point EXC, Measurement points IN1, IN2 and OUT, component transfer functions \textit{Piezo} (actuator), \textit{Table} (mechanical plant), \textit{Seis}mometer (sensor), \textit{Cont}rol filter and control switch SW.}
	\label{fig:TableTF}
\end{figure}

To design the active control system, we measured the transfer function of the table motion (measured with Nanometrics Trillium Compact 120s seismometer) against the piezo actuation. The measured and simulated transfer function is shown in Figure~\ref{fig:TableTF}.
Measurements generally agree with simulation other than the above-mentioned table resonant frequencies being lower than intended (Q factor and resonant frequency are fitted for the simulation curve) and an elevation in gain around the seismometer internal resonance at $\sim$0.008\,Hz.
The elevation in gain around the seismometer resonance is due to there being a tilt component in the horizontal piezo drive and the T120s being tilt-sensitive at low frequencies. 

Tilt coupling behaves like a geometric effect in which the table acts as a 100\,m tall pendulum. 
 For 20\,$\rm \mu$m of translation actuation there is 0.2\,$\rm \mu$rad of tilt induced in the table.
The tilt coupling mechanism is not understood.
This coupling has been estimated from the measured table transfer function shown in Figure \ref{fig:TableTF}.  
The purpose of the empirical model developed here is to show that the transfer function shape matches a tilt coupling model.
This transfer function can be divided into the sum of 2 transfer functions in two frequency bands.  
Below 0.1\,Hz the transfer function is dominated by the tilt coupling, sensed by the seismometer tilt to translation coupling. 
Above 0.1\,Hz the transfer function is dominated by pure translation.

Figure \ref{fig:TableTF} shows that the measured open loop transfer function (blue) does indeed match the tilt coupling theoretical model (dashed yellow) very well between 1mHz and 10Hz.
It is not consistent with a model without tilt coupling. 
The tilt coupling model explains some control limitations detailed below.

To determine the sensor noise floor, a 60\,dB gain 0.1\,Hz wide resonant gain filter centered at 0.6\,Hz was used as a controller.   
We determined the sensor noise limit to be 47.5\,dB below ground motion during a time of average seismic conditions for the site \cite{Satari_2022}. 
This was done by comparing the witness and in-loop seismometer.  The witness seismometer measured 47.5\,dB suppression as a ratio to the ground seismometer, while the in-loop seismometer measured 59.8\,dB suppression as a ratio to the ground seismometer.
This informed our target of 40\,dB suppression in the control band.

\begin{figure}[h!]
	\centering	\includegraphics[width=0.45\textwidth]{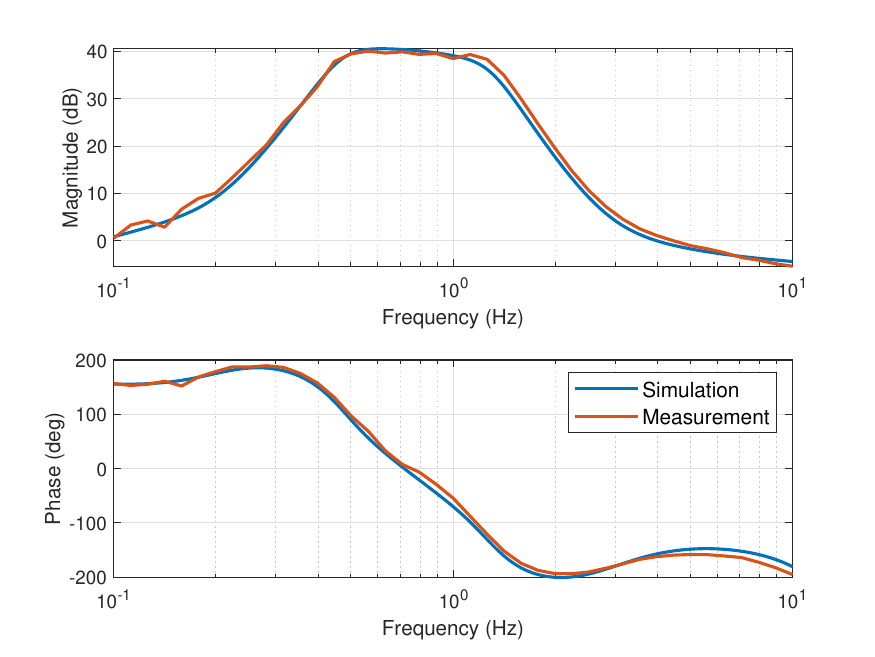}\\
    \caption{Open loop gain simulation and measurement for demonstration of broadband suppression of seismic noise.}
    \label{fig:ControllerOLG}
\end{figure}

\begin{figure}[h!]
	\centering
	\includegraphics[width=0.45\textwidth]{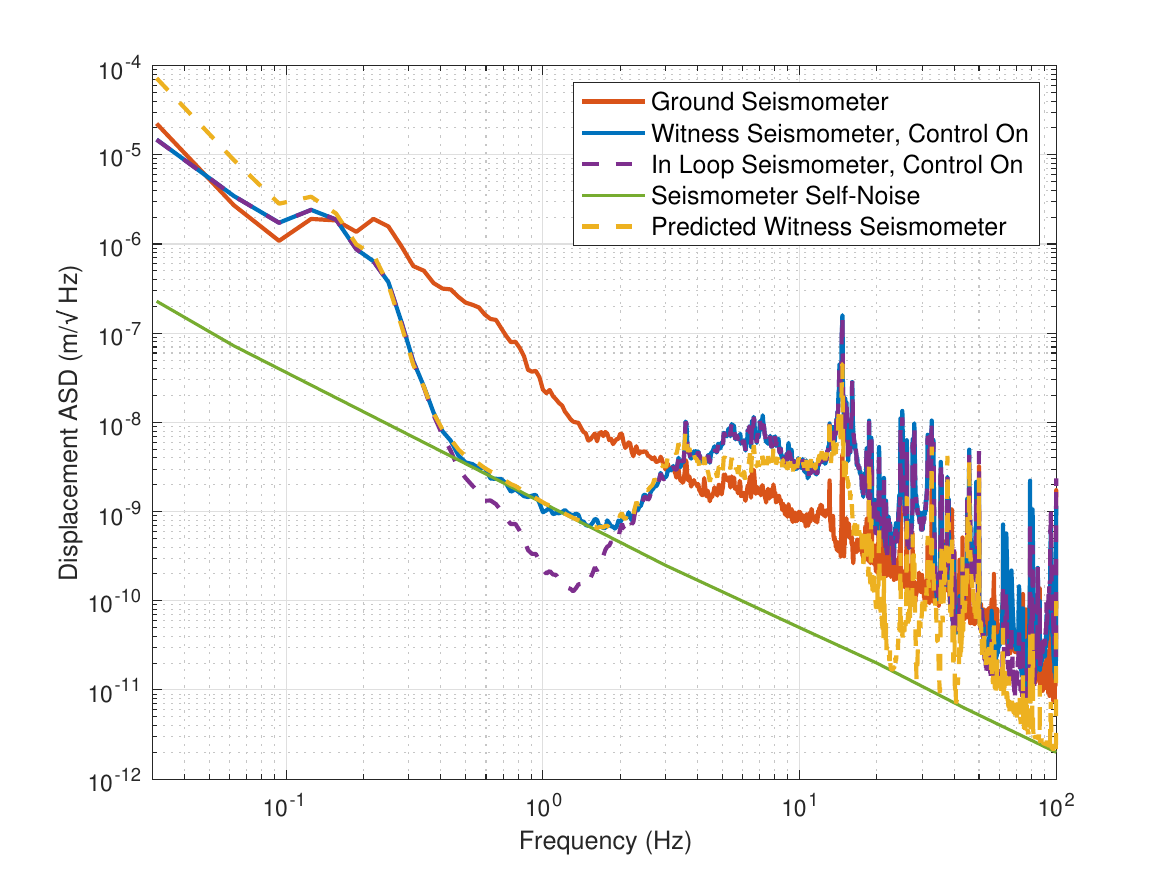}\\
	\caption{Amplitude spectral density (ASD) of the optical table and ground motion, with noise budget for predicted witness seismometer. The optical table motion is measured by two seismometers: one in loop (purple dash) and one witness seismometer (blue). Ground motion (red) is suppressed by the table in the control band. The fit to the measured seismometer self-noise is shown in green. Elevated witness seismometer relative to in loop seismometer is roughly predicted by seismometer self-noise around 1\,Hz. This predicted witness seismometer noise (yellow dash) is the quadrature sum of Self-Noise and the predicted seismometer noise, calculated as Ground $\times$ Table $\times$ 1/(1+G) where G is the open loop gain in Figure 7 and Table is the transfer function in Figure 6.}
	\label{fig:TableSupression_ASD}
\end{figure}

We therefore design a 'butter-worth band-pass boost filter' as an 40\,dB gain AC-coupled controller.  
The filter is the ratio of 2 3$^{rd}$ order Butterworth band-pass filters with cutoff frequencies (0.2,3)\,Hz and (0.4,1.4)\,Hz.    
This is similar to a resonant gain filter that is composed as the ratio of pairs of complex conjugate poles (resonances) that have the same resonant frequency but different Q factor.
The open loop gain measurement and model of the system is shown in Figure \ref{fig:ControllerOLG} measured as IN2/IN1 with swept sine excitation EXC and switch SW open in the lower panel Figure~\ref{fig:TableTF}, control and measurement is done with the LIGO Control and Data System~\cite{CDS}. 
Displacement spectral density measurements of the inertial vibration isolated table and ground are shown in Figure \ref{fig:TableSupression_ASD}. 
Comparing ASD from ground and witness seismometer with and without control, suppression of the optical-table motion can be observed from 0.16 to 3\,Hz.
The maximum suppression of 36\,dB is observed around 0.5\,Hz, with suppression levels ranging from 21 to 36\,dB within the frequency range of 0.5 to 1.3\,Hz.

The suppression reduces the residual horizontal motion ASD from 20\,$\mathrm{nm/\sqrt{Hz}}$ to 1\,$\mathrm{nm/\sqrt{Hz}}$  at 1\,Hz.
There is some amplification of signals outside the control band which can be attributed to insufficient phase margin and gain margin in the control system. 
These limitations result in a trade-off between isolation strength, bandwidth and system stability for this controller, where improving isolation strength and bandwidth, weakens stability at the band’s edges.
As we target reducing table motion at suspension resonant frequencies, we have accepted increased table motion at higher frequencies until a more optimal controller is designed.
The effect of tilt-translation coupling of the seismometer complicates control at frequencies below 0.1\,Hz.
This is the result of tilt creating larger signals than translation.
If the control band is lowered, these signals exceed the dynamic range of the actuator and make the controller unstable.  
We were able to reduce the actuation range by using an actuation matrix that subtracted tilt with the vertical actuators, using the tilt coupling model.
The time variation of tilt coupling limited the degree of tilt decoupling that could be achieved.

The difference between witness and in-loop seismometer between 0.4 and 2\,Hz indicates control is sensor noise limited in this band. 
Self-Noise depicted in Figure \ref{fig:TableSupression_ASD} is a fit to a measurement of the residual between the in-loop and witness seismometer from a time where no control was applied (known as a huddle test).  It is almost a factor of 10 higher than the instrument specification at 1\,Hz, indicating the potential for improvement.

\begin{figure}[h!]
 	\centering
 	\includegraphics[width=0.45\textwidth]{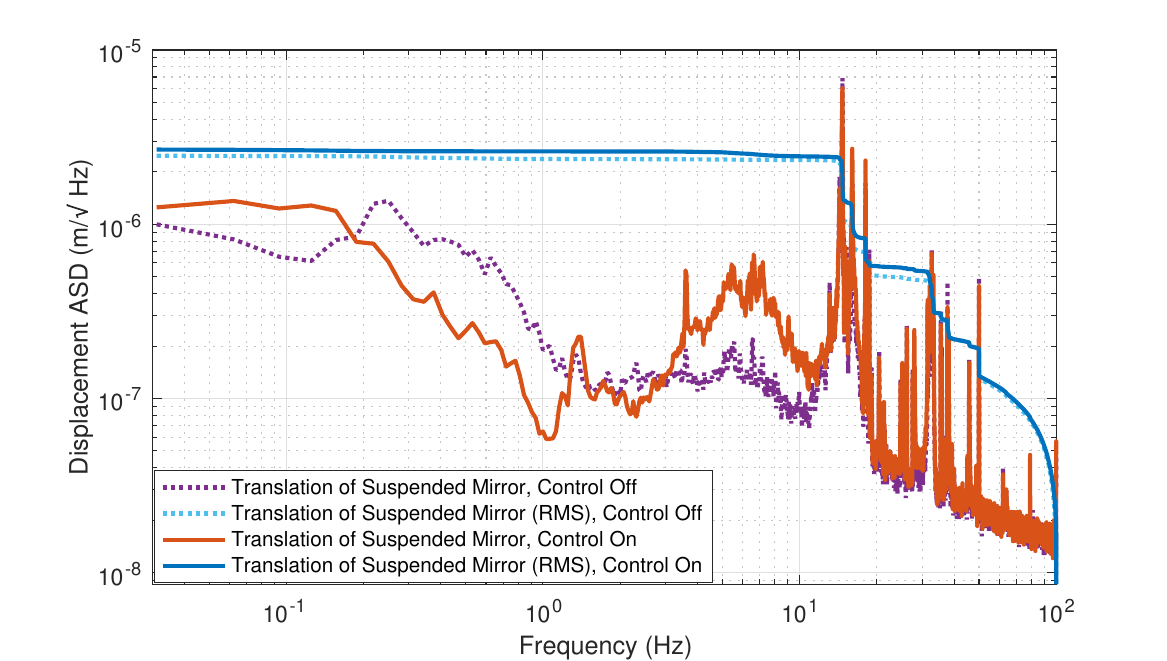}\\
 	\caption{Cavity length motion of the suspended test mass relative to the supporting structure that is attached to the vibration isolated table.  Reduction (red solid vs purple dash) in residual relative motion is observed in the control band. Root mean squared relative motion (blue solid and dash) is dominated by table resonances.}
 	\label{fig:SusSupression}
 \end{figure}

There is a suspended test mass on the table that uses an initial LIGO suspension \cite{InitialLIGO} with OSEM \cite{abbott2009advanced} comprised of an opto-electronic sensor and coil magnet actuator.
The suspended test mass has damping loops to suppress the suspension resonant motion. 
Ideally, we would see a reduction in relative motion of the test mass above the suspension resonance if the motion was simply excited by seismic noise.  
What we see in Figure \ref{fig:SusSupression} is there is a factor of 2 reduction in broadband relative motion between the table and the test mass on the table suppression band and there is a small increase in resonant motion at 1.5\,Hz.

This implies that seismic noise is not the limiting noise in this measurement and that the table control has interaction with the suspension, increasing motion on resonance.
As the vacuum chamber has not yet been evacuated acoustic noise could be responsible for the elevated motion.

\section{Conclusion}\label{sec:performance}
Tilt coupling results in large-low frequency error signals that surpass the dynamic range of the piezo actuator below 100\,mHz.  
The simple controller used has poor phase margin at high frequencies, this is compounded by table resonant frequencies being lower than expected adding extra phase delay. Currently this combination limits table control to below a few Hz. 

The resonant frequencies have been observed to change with changes in the foot height adjustment.  
This could potentially provide a means to tune the pressure on each of the four-footed table feet.
It would be possible to compensate the resonances in the control filter to achieve higher frequency control. 
The table resonances have a Q factor of approximately 70.  This Q factor would require finely tuned compensation.
The plan is to actively and passively damp the resonances first and then compensate them in control.

To correct the tilt coupling we plan to fix the pressure on the feet to reduce inherent coupling.  
We will also use the tilt correction system in the future by applying an actuation matrix for translation comprised of translation and tilt. 
Correction of tilt coupling would allow table isolation over more of the microseism frequency band.

We have demonstrated that an inertial isolated piezo actuated table is capable of 21 to 36\,dB suppression of ground motion within the frequency ranging from 0.5 to 1.3\,Hz. The suppression results in 1\,$\mathrm{nm/\sqrt{Hz}}$ residual horizontal motion at 1\,Hz.  
We have described the issues of tilt coupling and internal resonances present in the table design and proposed mechanisms to overcome these issues.
This would allow a wider frequency band to be controlled with sensor-limited noise performance.

\section*{Acknowledgements}
We wish to thank the LIGO Scientific Collaboration Suspensions Working Group for valuable discussion. This work is supported by the Australian Research Council Center of Excellence for Gravitational Wave Discovery project CE170100004. 

The authors would like to thank Felix Wojcik and John Moore who contribute so much to the maintenance and technical improvements made to the Gingin site. 
Appreciation is also extended to Bram Slagmolen for assistance with the CDS and to Bastien Lassalle for support in preparing the assemblies for vacuum.

\bibliography{refs}
\end{document}